\documentclass{PoS}

\title{Jet Production Measurements at CMS}

\ShortTitle{Jet Production Measurements at CMS}

\author{\speaker{Sanmay Ganguly}\\
        Department of High Energy Physics\\
        Tata Institute of Fundamental Research\\
        Mumbai, India \\
        E-mail: \email{Sanmay.Ganguly@cern.ch}}

\abstract{Jet production cross-section measurements are presented. The measurements are done with the data
          from Large Hadron Collider (LHC) proton-proton collisions, collected with the Compact Muon Solenoid (CMS) detector.
          The inclusive jet production measurements are carried out with data collected at 
          $\rm \sqrt{s}~ =~ 7~ TeV$ and $\rm 8~TeV$ with total integrated luminosity ($\mathcal{L}_{int}$)
          $\rm 5.0~ fb^{-1}$ and $\rm 10.71~ fb^{-1}$ respectively. The dijet production measurements are carried out with
          the $\rm \sqrt{s}~ =~ 7~ TeV$ dataset. Jets are reconstructed with the anti-$k_T$ clustering algorithm with size parameter
          $R=0.7$. The measured cross sections are corrected for detector
          effects and compared to perturbative QCD predictions at NLO, corrected for
          NP factors, using various sets of PDF.  
          The inclusive jet cross-section ratio of the jets reconstructed with the anti-$k_T$ (AK) algorithm and two radius parameter $\rm R~=~0.5$ and $\rm R~=~0.7$
          are also presented. The data used is $\rm \sqrt{s}~ =~ 7~ TeV$ CMS data corresponding to $\rm \mathcal{L}_{int}~=~5.0~ fb^{-1}$.  
          Significant discrepancies are found comparing the data to leading order calculations and to fixed order calculations at NLO, corrected for NP effects,
          whereas simulations with NLO matrix elements matched to the parton showers describe the data quite well. 
          A study of color coherence effects in pp collisions has been performed with the data collected at $\rm \sqrt{s}~ =~ 7~ TeV$ and $\rm\mathcal{L}_{int}~=~ 36~ pb^{-1}$. 
          The measurement of the azimuthal angular correlation between the second and third jets is compared to the predictions of Monte Carlo models
          with different implementations of color coherence effects.
          }

\FullConference{The European Physical Society Conference on High Energy Physics -EPS-HEP2013\\
                18-24 July 2013\\
                Stockholm, Sweden}

\begin{document}

\section{Introduction}
Quantum Chromo Dynamics (QCD) is a gauge quantum field theory which describes the interaction among sub-nuclear colored partons, viz. quarks, gluons etc. The theory predicts scattering 
cross-section of partons for a fixed order of perturbation theory. In a multi-hadron collision scenario at high center of mass (COM) energy, the fundamental
 interacting particles are partons whose initial
four-momenta are not fixed and their distribution is described by parton distribution function (PDF). The outgoing scattered partons fragment and hadronize to form colorless stable particles, 
viz. hadrons. The fixed order perturbative QCD (pQCD) calculations along with non-perturbative (NP) hadronization model and PDF parametrization, predicts the dynamics and cross-section of a hadron 
collision processes within uncertainty limits.

 The QCD theory suffers from divergences originating from soft and collinear branching of the partons. In order to remove these divergences from the observables computed out of this theory,
certain clustering algorithm are designed to cluster the outgoing colorless hadrons. The clustered particles are combinedly called jet whose four momenta carries a signature of parton
level kinematics in the collision phenomena. By virtue of these recombination algorithms, the jet observables are free from any kind of divergence, making them a key variable of study in experimental
measurements. 

 At LHC proton-proton collisions take place at high COM energy ($\sim$ few TeV). At LHC energy scale the fundamental colliding objects are gluons and quarks and hence QCD is the most 
dominant interaction process with very large cross-section. In order to test and validate QCD theory predictions at LHC, jet observable measurements become an important and interesting study. 
These measurements at LHC help to test and validate QCD predictions at unexplored kinematic regimes. Using the jet variables one can optimize the hadronization and parton shower
models and also can constrain the parameters of PDF's. The running of strong coupling constant $\alpha_{S}$ in the new kinematic regimes can also be extracted using the jet observables. QCD processes 
act as a major background to other standard model process and new physics searches. A precise estimation of QCD background becomes essential in those studies, which can be accurately done using 
jet observables. 

\section{Jet reconstruction and jet energy correction}
In CMS detector at LHC \cite{DETECTOR}, several event reconstruction techniques are used and one of the most popular among them is the  paricle flow (PF) technique. PF algorithm \cite{PFLOW}
 is an event reconstruction technique 
which attempts to reconstruct and identify all stable particles in an event from all sub-detector components  viz. tracker, electromagnetic calorimeter (ECAL), hadronic calorimeter (HCAL), muon detector.
In a typical collision scenario, the stable particles reconstructed are $\mu^{\pm},~~ e^{\pm},~~ \gamma,~~ \pi^{\pm},~~ \pi^{0},~~ K^{\pm},~~ K^{0}$, which along
with neutron and proton are reconstructed by jet clustering algorithm to form jets. 

\begin{figure}
\includegraphics[scale=0.23]{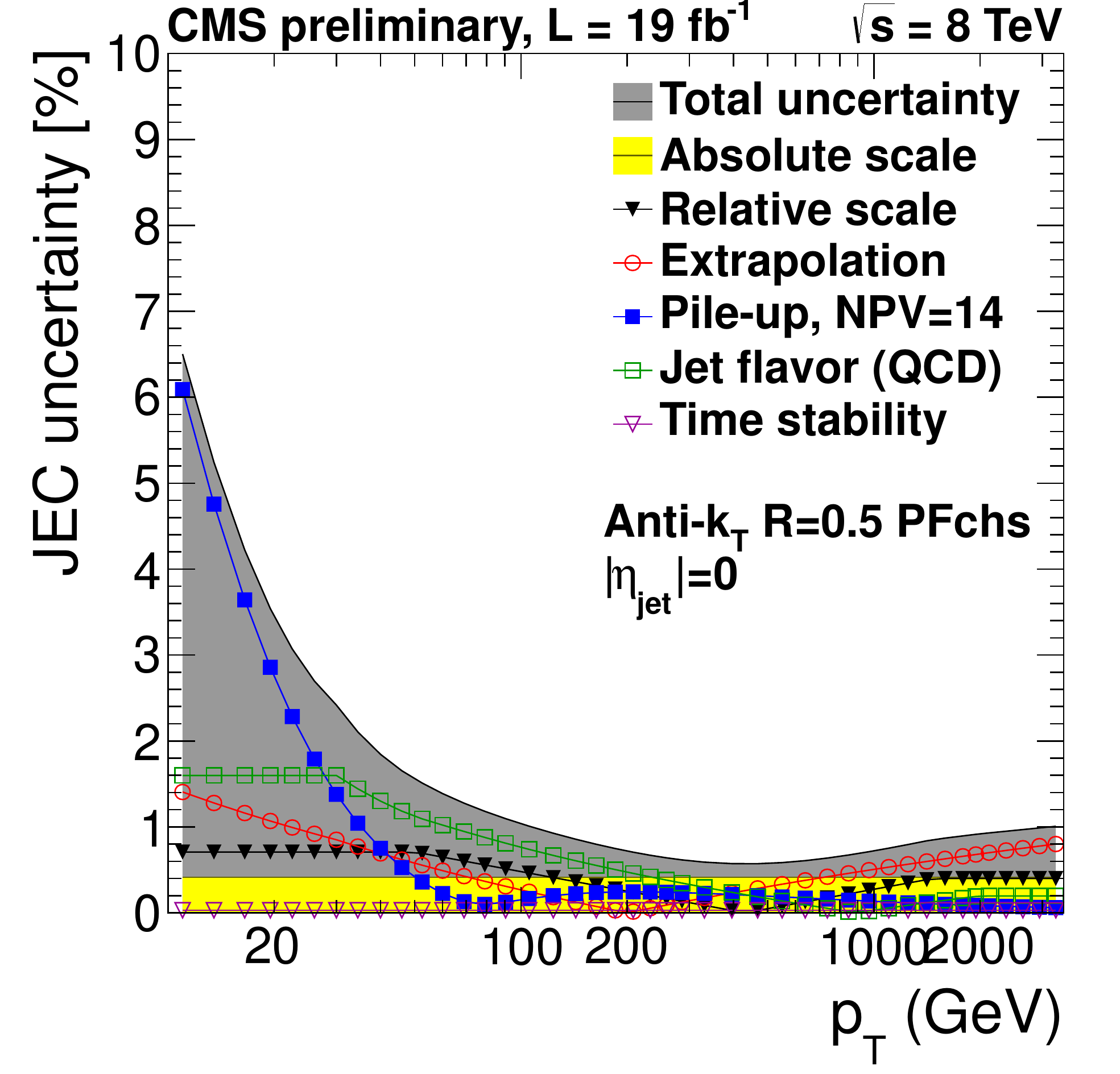}
\includegraphics[scale=0.23]{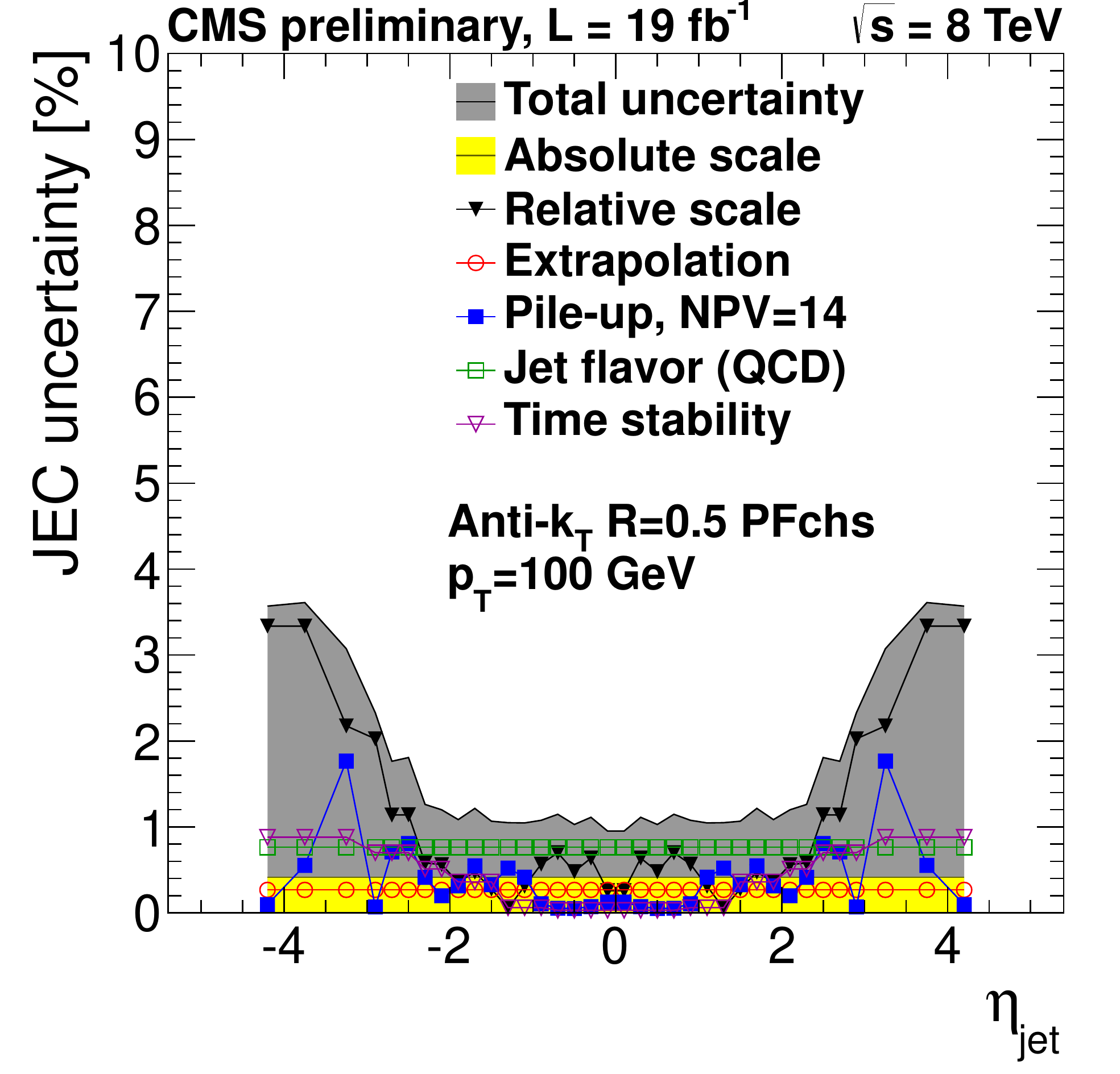}
\includegraphics[scale=0.43]{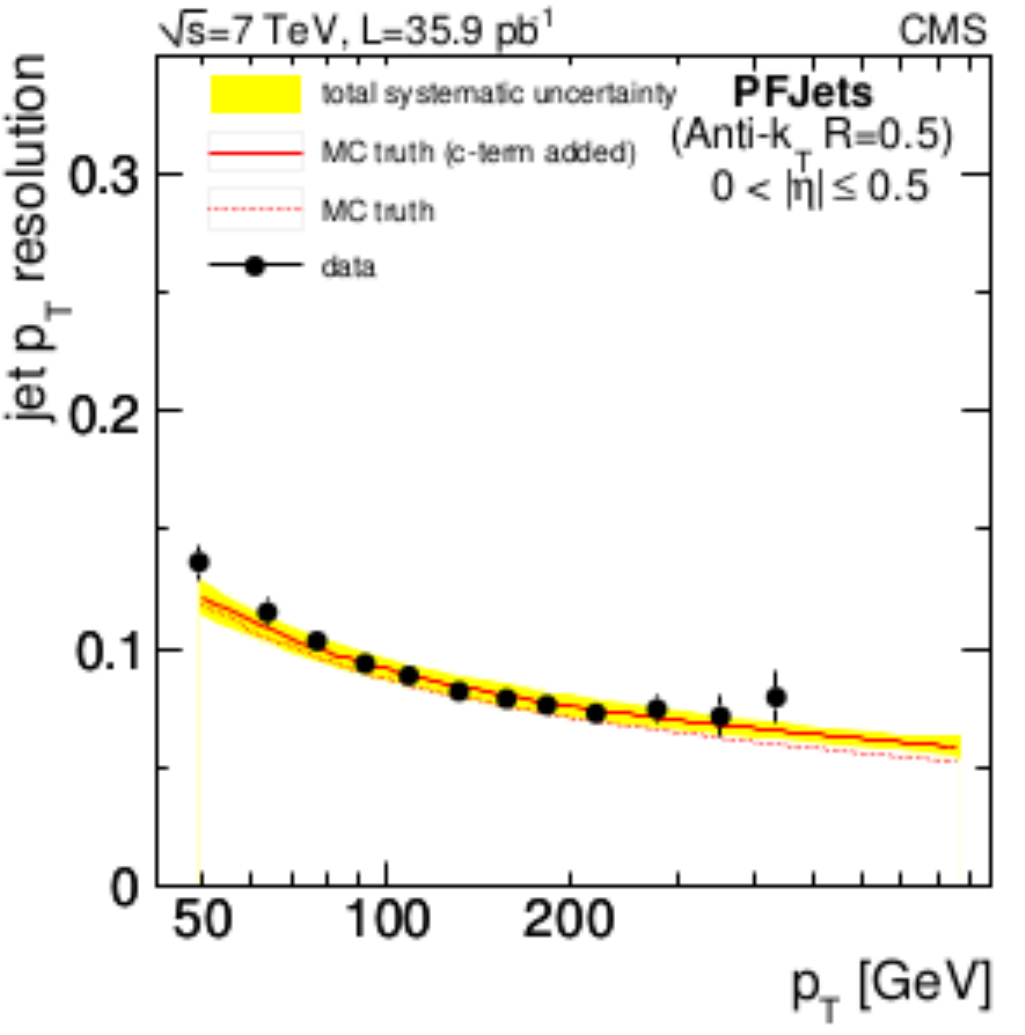}
\caption{The left plot shows the JEC uncertainty on the measured four momenta due to different individual sources, for the central jet, as a function of jet $\rm p_{T}$. The 
 middle plot shows the JEC uncertainty as a function of $\rm \eta_{jet}$ for each sources in high $\rm p_{T}$ region. For both the plots, the total uncertainty (grey band) is a quadrature some of the 
individual sources. The right plot shows the jet energy resolution as a function of jet $\rm p_{T}$ for central jets derived both from data and MC.}
\label{fig:uncsrc}
\end{figure}

One of the widely used jet reconstruction algorithm in CMS is anti-$\rm k_{T}$ (AK) algorithm \cite{AKT} with $\rm R~=~0.5$ and $\rm R~=~0.7$. The reconstructed jet four momenta 
are corrected for the additional energy reconstructed, originating from multiple proton-proton scattering vertices in a single bunch crossing, called  pileup. The jet four momenta
are further corrected to compensate for the non uniform detector response 
along jet pseudo-rapidity ($\rm \eta_{jet}$), viz. residual correction and jet transverse momentum ($\rm p_{T}$) termed as absolute correction.
 An additional correction is also imposed to account for data-MC mismatch,
 called residual correction.
In Fig.\ref{fig:uncsrc} the uncertainty on the jet energy scale (JES), due to jet energy correction (JEC) is shown in the first two plots \cite{JESNOTE}.
 With the new derived sources, the total JES uncertainty
goes down below $1\%$ for the central jets around jet $\rm p_{T}~=~500~GeV$. The same is true for high $\rm p_{T}$ in central region. The right most plot shows the jet energy resolution as a
 function of jet $\rm p_{T}$ for central jets, which is about $9\%$ for jet $\rm p_{T}~ \sim ~100~GeV$.

\section{Jet cross-section measurements}
At CMS jet production double differential cross-sections are measured as a function of
 jet $\rm p_{T}$ and dijet invariant mass ($\rm M_{jj}$) for different rapidity bins, respectively \cite{JET7,JET8}.
 The jets are selected with tight jet identification criteria, to reject detector level noise,
 followed by JEC applied on selected jets. The double differential cross section are measured using the formula
 \begin{equation}
 \rm  
 \frac{d^2 \sigma }{dp_{T}dy} = \frac{1}{\epsilon \mathcal{L}_{int}}\frac{N}{\Delta p_{T}~ (2~\Delta |y|)} \times C_{unsmearing} 
 \end{equation}
 \begin{equation}
 \rm
 \frac{d^2 \sigma}{dM_{jj}dy} = \frac{1}{\epsilon \mathcal{L}_{int}}\frac{N}{\Delta M_{jj}~ (2~\Delta |y|)} \times C_{unsmearing}
 \end{equation}
 where $\mathcal{L}_{int}$ is the total integrated luminosity, $\epsilon$ is the trigger efficiency, $N$ is the number of jets in a bin.
 $\rm \Delta p_{T}$, $\rm \Delta M_{jj}$, $\rm \Delta |y|$ are the inclusive jet $\rm p_{T}$, $\rm M_{jj}$ and jet rapidity bin widths respectively and $\rm C_{unsmearing}$ is
 the unfolding correction factor.

\begin{figure}
\hspace{0.5cm}
\includegraphics[scale=0.27]{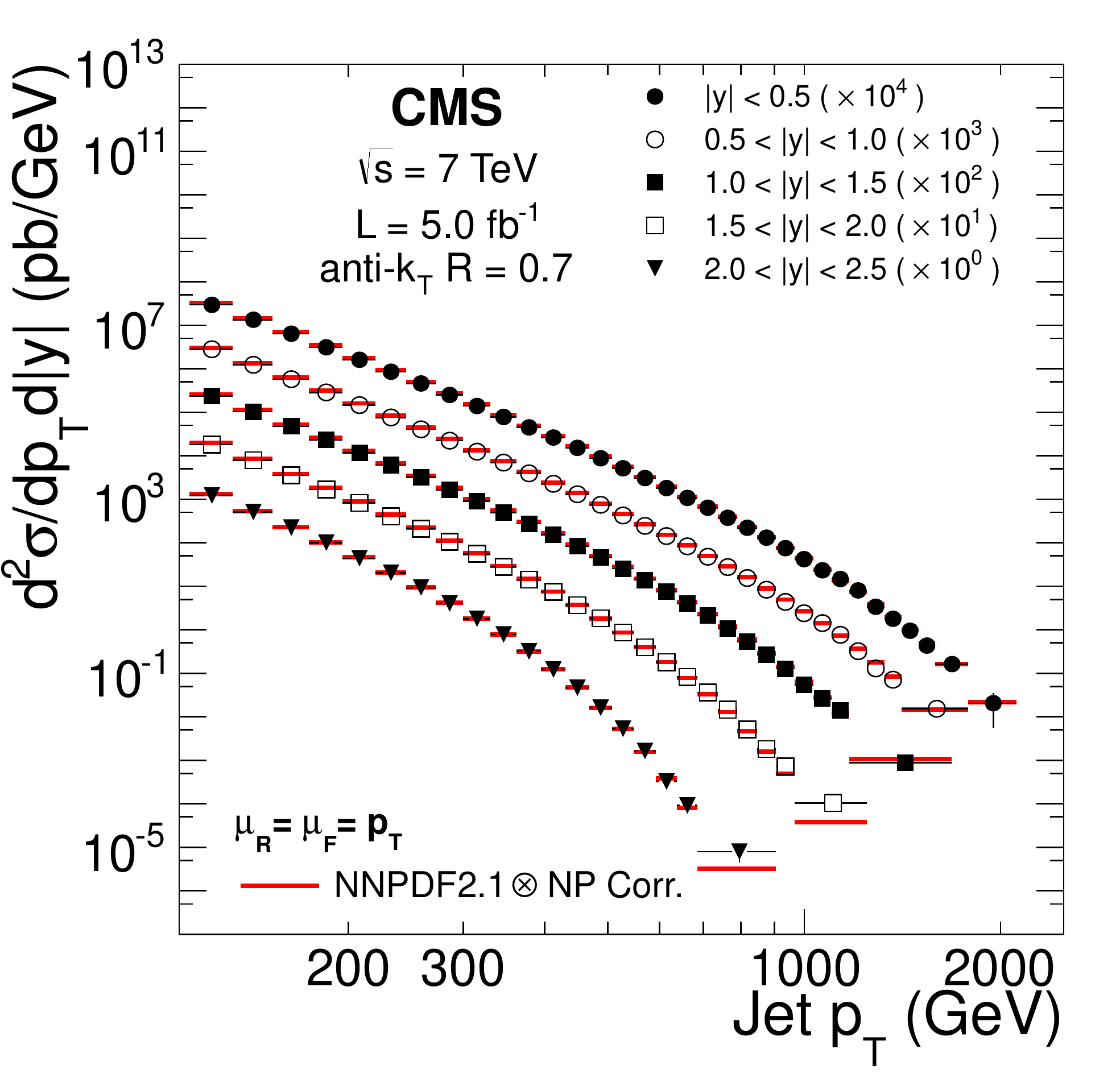}
\hspace{1.5cm}
\includegraphics[scale=0.27]{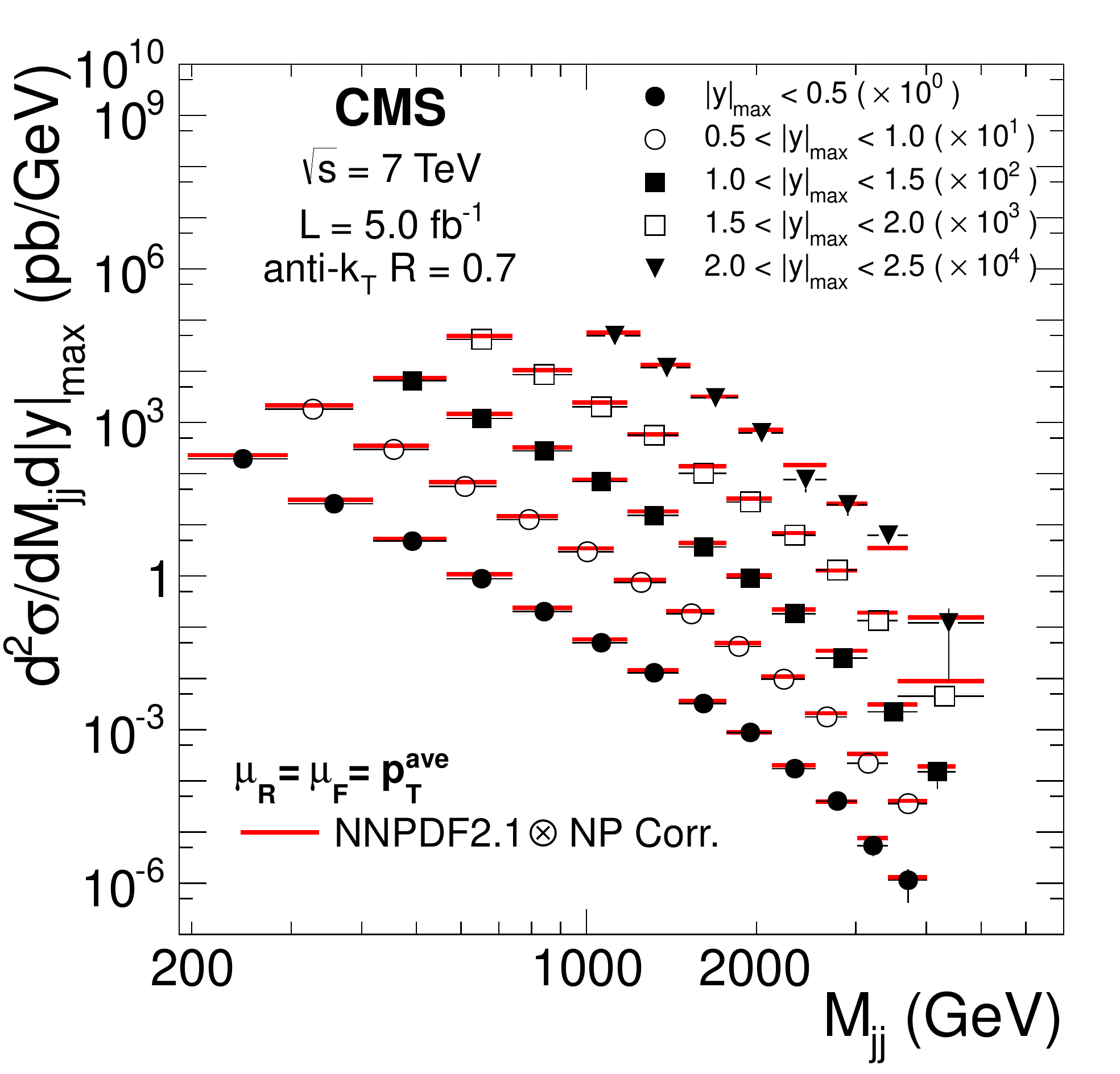}
\caption{The left plot shows the measured double differential inclusive jet production cross-section (black points) for five different rapidity bins at an interval of $\rm \Delta |y|~=~0.5$
 extending upto $\rm |y|~=~2.5$. The measured unfolded cross-section is compared with $\rm NLO \otimes NP$ theory prediction (red line), calculated with NNPDF2.1 PDF set.}
\label{fig:spectrum}
\end{figure}

 For the data collected at  $\rm \sqrt{s}~ =~ 7~ TeV$ ($\rm \mathcal{L}_{int}~=~5~fb^{-1}$) with CMS detector, the inclusive jet cross-section measurements extend upto 
jet $\rm p_{T}~=~2~TeV$ and rapidity $\rm |y|~=~2.5$ with an interval of $\rm \Delta |y|~=~0.5$. The dijet invariant mass measurement extends upto $\rm M_{jj}~=~5~TeV$ with
 the similar rapidity coverage as inclusive jet measurements as shown in Fig.\ref{fig:spectrum}.

 For the inclusive
 jet measurement, events are required to contain at least one tight jet with $\rm p_{T}>$114 GeV, 196 GeV, 300 GeV,
 362 GeV, and 507 GeV for the five single-jet high level triggers (HLT)  used respectively. For the dijet measurement, at least two tight
 reconstructed jets with $\rm {p_{T}}_1>$ 60 GeV and $\rm {p_{T}}_2>$ 30 GeV are required. The measured spectra are unfolded using Bayesian unfolding (D'Agostini)\cite{ROOUNFOLD} technique 
 to remove detector smearing effects. The measured cross-section is then compared to NLO theory predictions corrected by the NP correction factor. 
 The later is required in order to account for the hadronization and multi-parton interaction (MPI) effects.

 The next to leading (NLO) theory calculation is done by NLOJet++ package using fastNLO program \cite{FNLO} using five different PDF sets viz. ABKM09, MSTW2008, HERA1.5, CT10, NNPDF2.1.
 The renormalization scale ($\rm \mu_{R}$) and factorization scale ($\rm \mu_{F}$) set for the NLO
 calculation are the jet $\rm p_{T}$ and $\rm p_{T}^{avg}~=~ \frac{1}{2}({p_{T}}_1 + {p_{T}}_2)$ for the inclusive and dijet production cross-section calculations, respectively.
 The scale uncertainty due to fixed order theory calculation is estimated by varying the $\rm \mu_{R}$ and $\rm \mu_{F}$ for the six points
 $\rm \frac{\mu_{F}}{\mu},\frac{\mu_{R}}{\mu}=(2,2),(0.5,0.5),(1,0.5),(0.5,1),(2,1),(1,2)$, where $\mu$ is kept fixed to jet $\rm p_{T}$ or $\rm p_{T}^{avg}$ for the inclusive and 
 dijet cross-section calculations respectively. The NP factor is derived by taking 
 the ratios of two cross-section predictions, calculated by tuning on and off the hadronization and MPI effects. For this analysis, the NP correction factor is derived using PYTHIA6 (uses Lund string
fragmentation model) and HERWIG++ (uses cluster fragmentation model)
 generators and difference between these two predictions are quoted as the uncertainty in the NP factor.  
 
\begin{figure}
\includegraphics[scale=0.27]{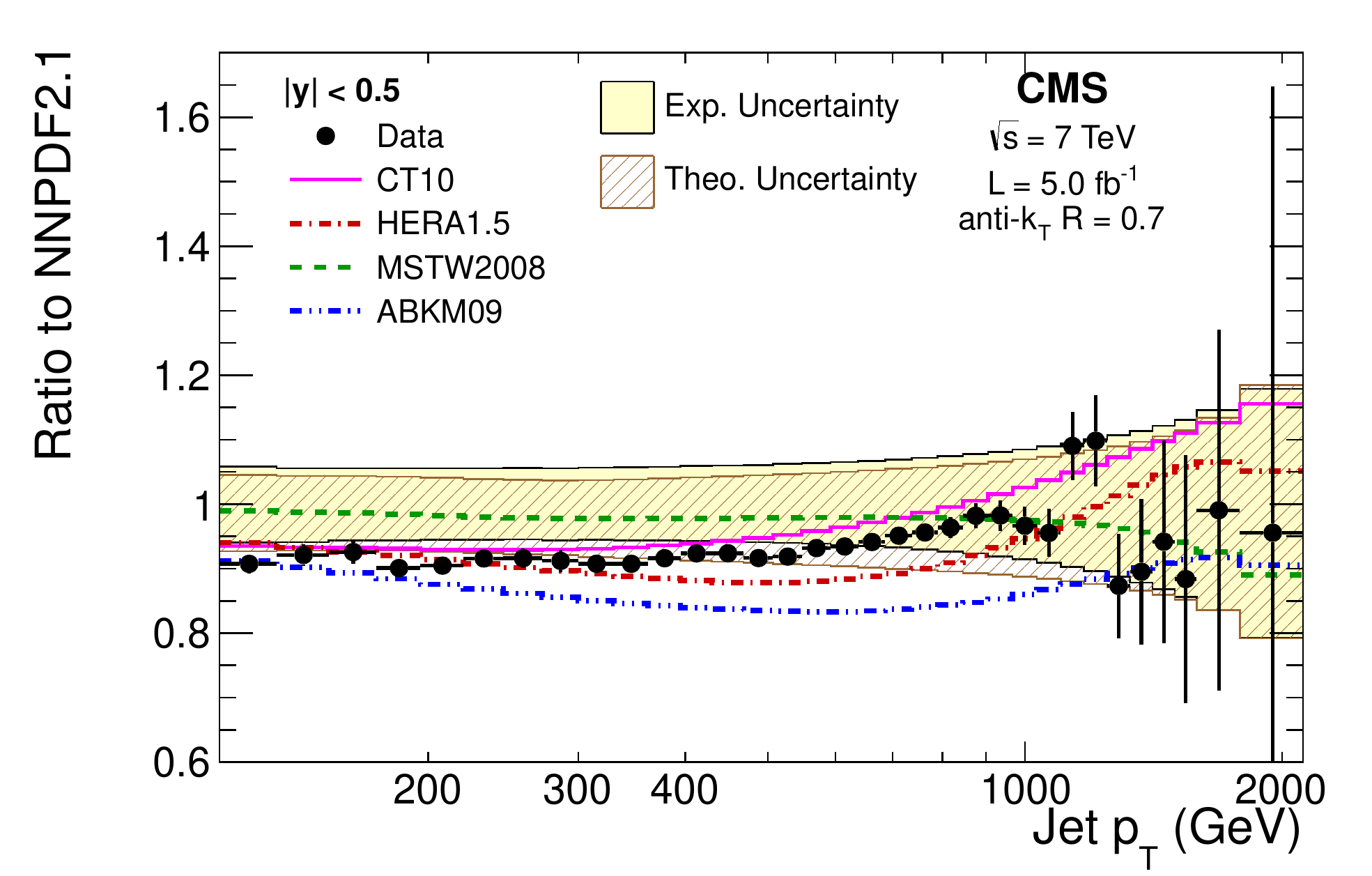} \hspace{1.5cm}
\includegraphics[scale=0.27]{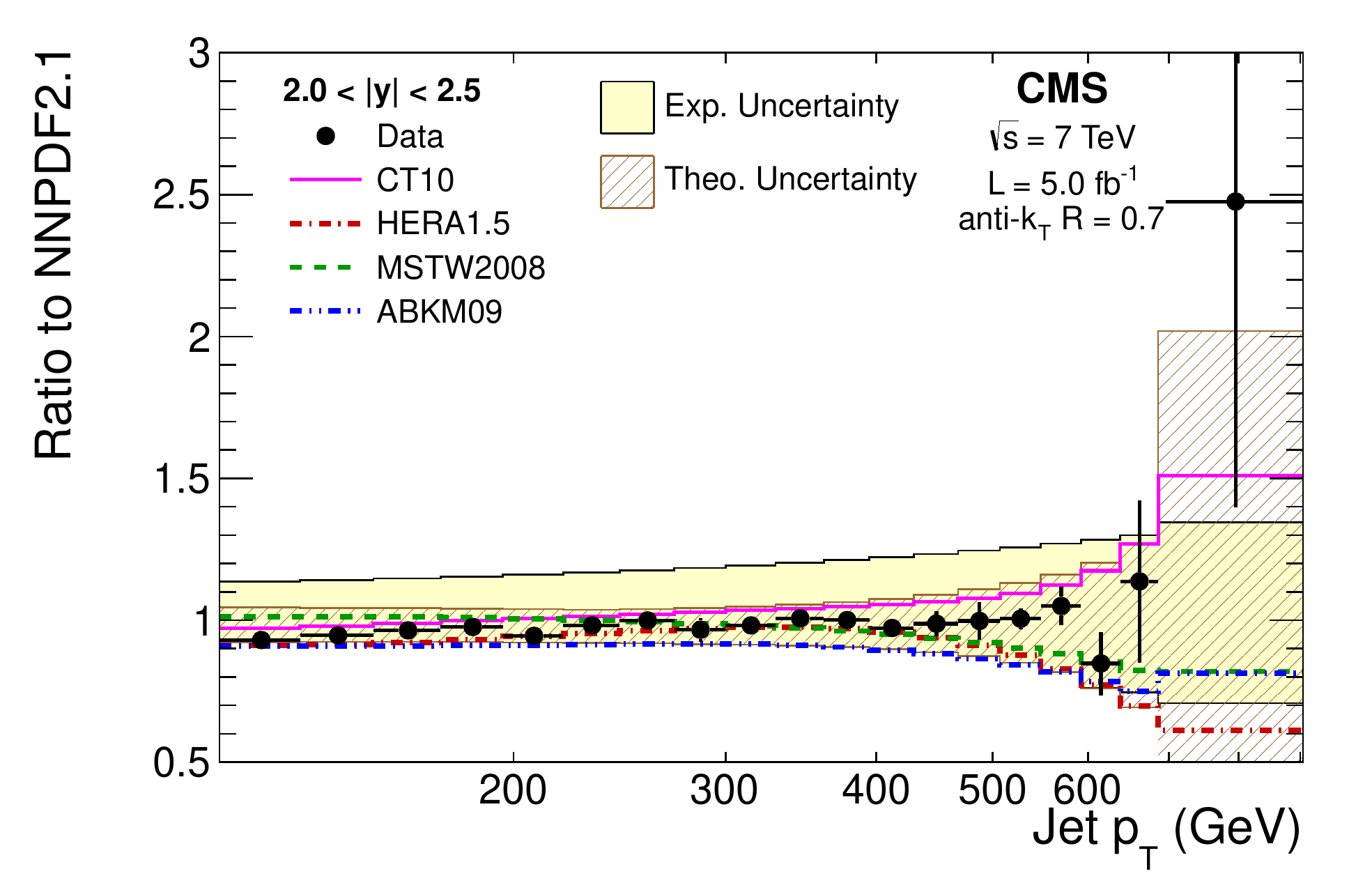}\\
\includegraphics[scale=0.27]{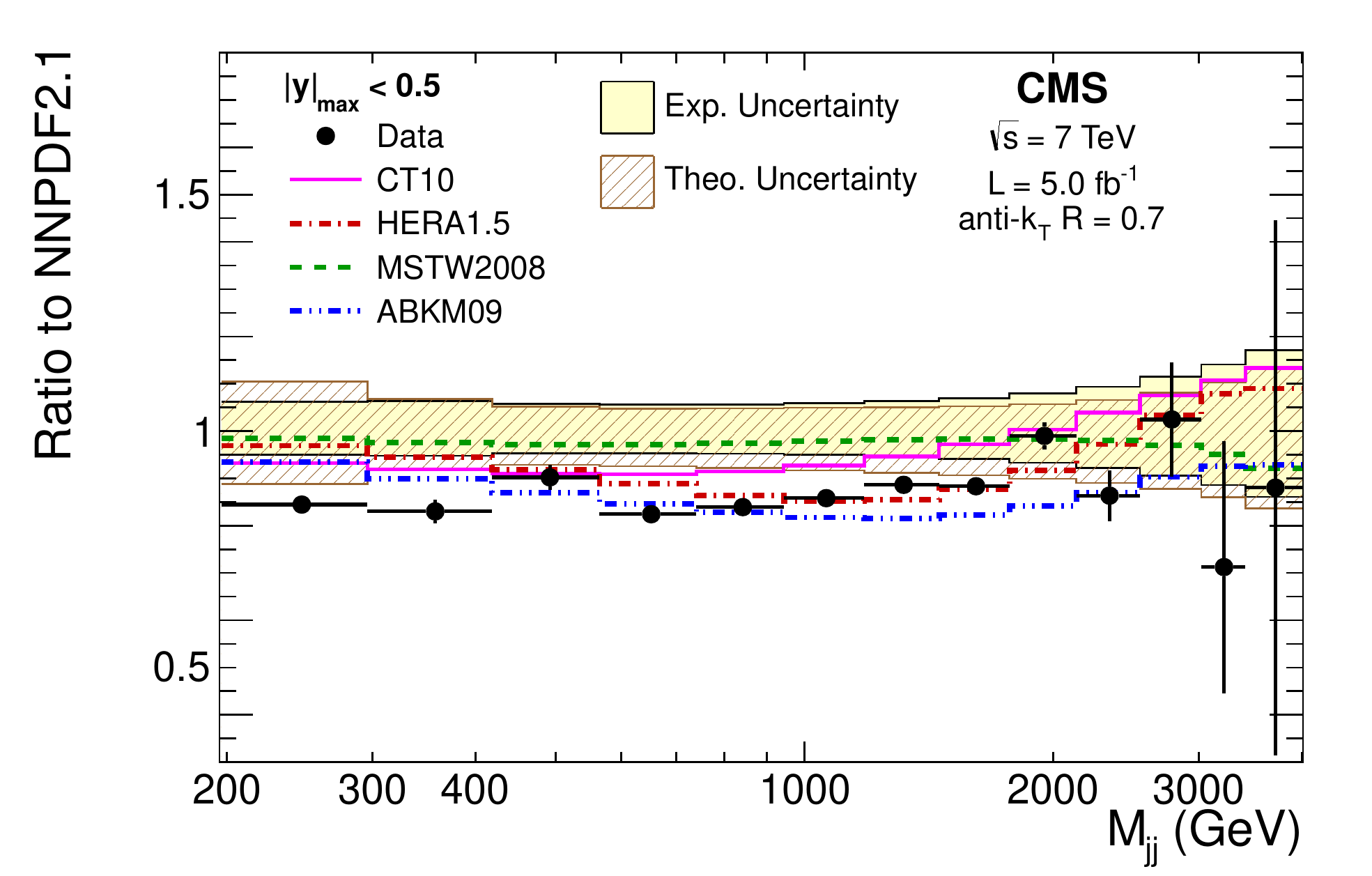}    \hspace{1.5cm}
\includegraphics[scale=0.27]{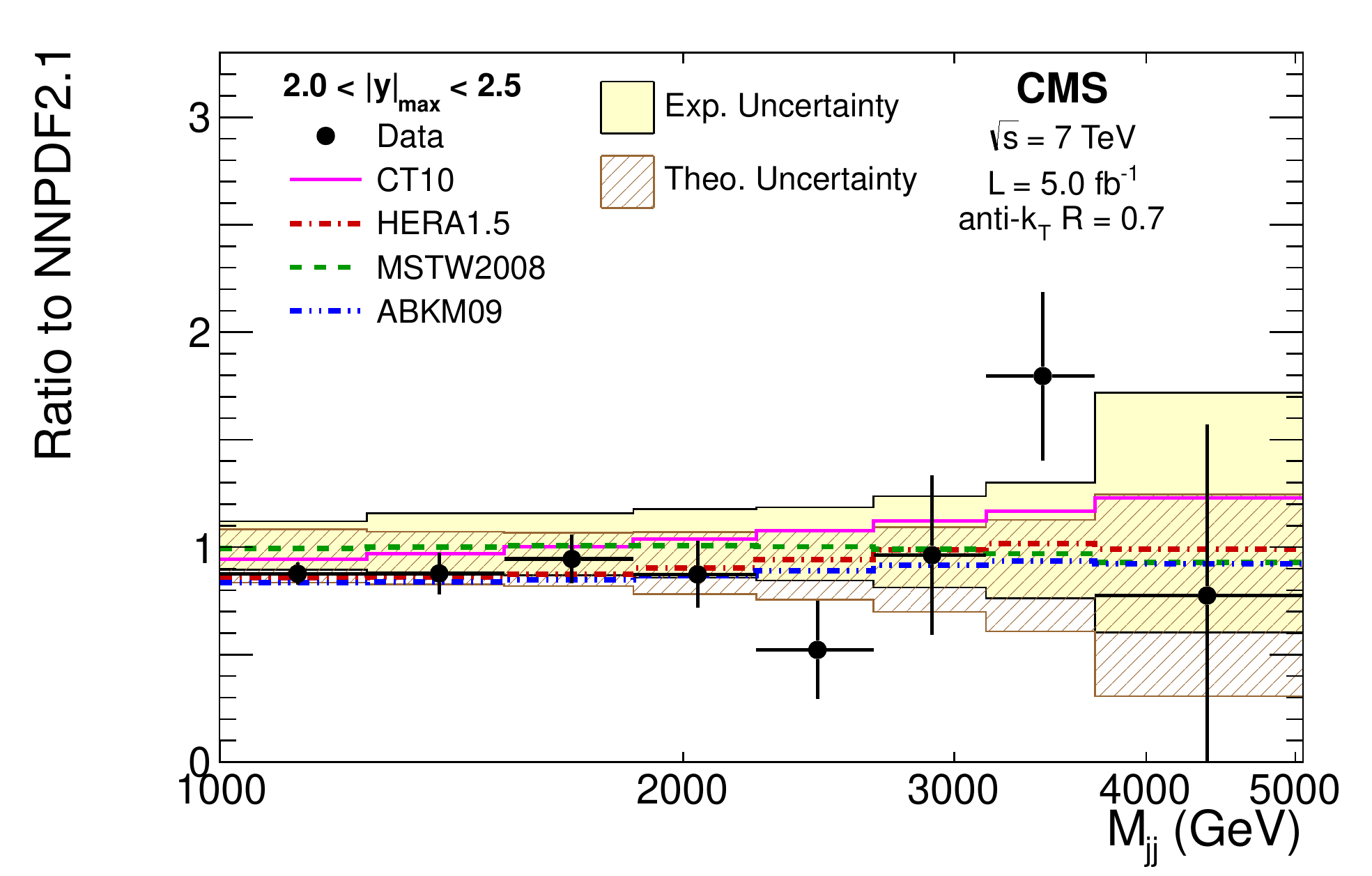}
\caption{The ratio of measured cross-section to the NNPDF2.1 theory prediction is shown along with the ratio to other PDF sets and relative total experimental and theoretical uncertainty for two extreme 
rapidity regions for inclusive (top) and dijet (bottom) spectrum. In general a good agreement between data and theory is observed.}
\label{fig:ratio7}
\end{figure}
 
In Fig.\ref{fig:ratio7} we see that in the central rapidity region ($0<|y|<0.5$), data is within 5-10$\%$ agreement with the NNPDF2.1 theory prediction.
 The CT10 prediction fluctuates up-to 20$\%$ in this $\rm |y|$ bin
and the total theoretical and experimental uncertainties are within 5-20$\%$ limit for the inclusive jet case. For the outer $\rm |y|$ bin ($2.0<|y|<2.5$) the experimental uncertainty
 reaches up-to 40$\%$. For the 
dijet measurement the data is within 20$\%$ agreement with NNPDF2.1 theory for central region where as deviates up-to 80$\%$ in the outer $\rm |y|$ bin where experimental uncertainty also increases
to 60$\%$.

\begin{figure}
\hspace{0.5cm}
\includegraphics[scale=0.28]{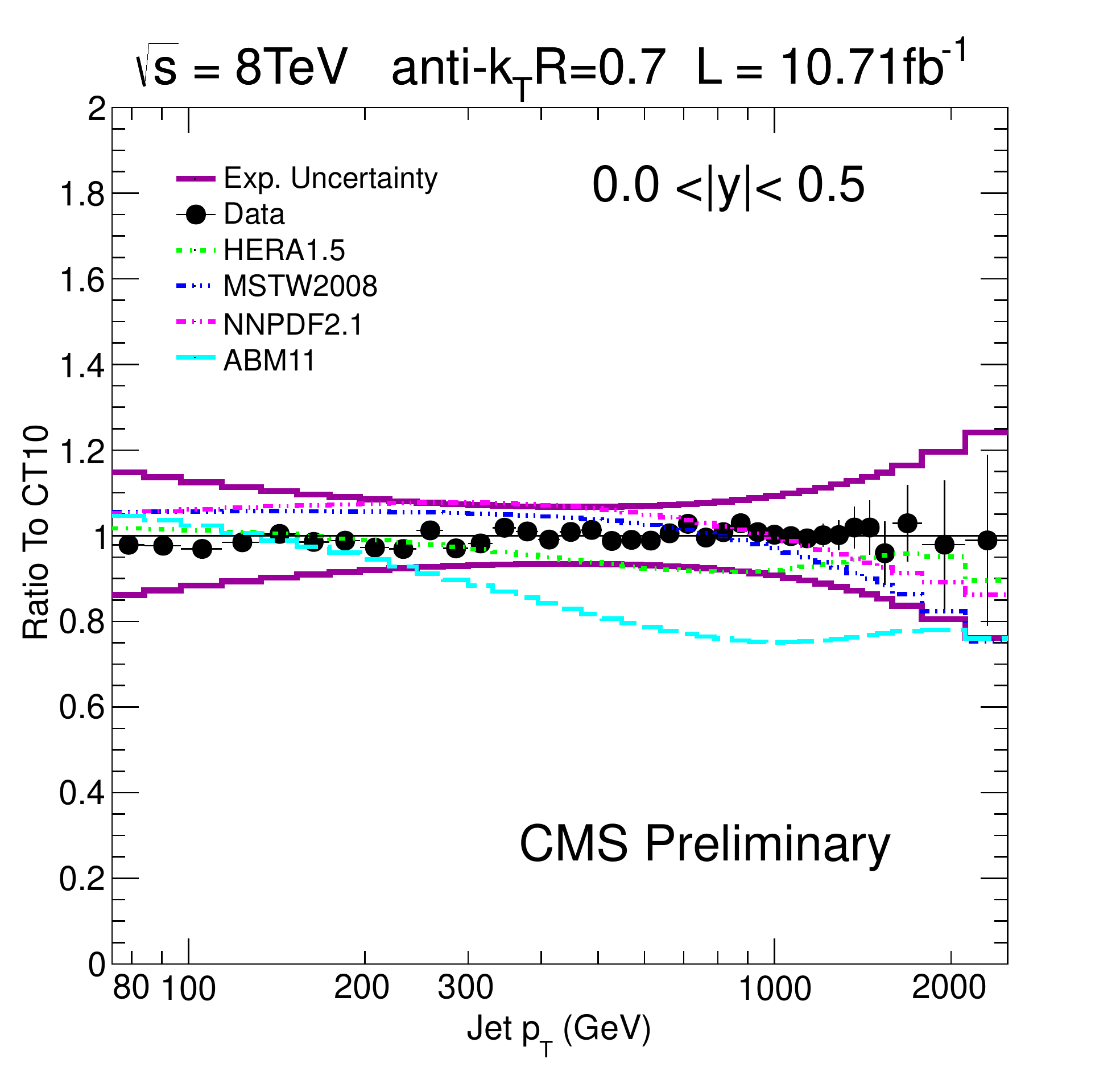} \hspace{1.5cm}
\includegraphics[scale=0.28]{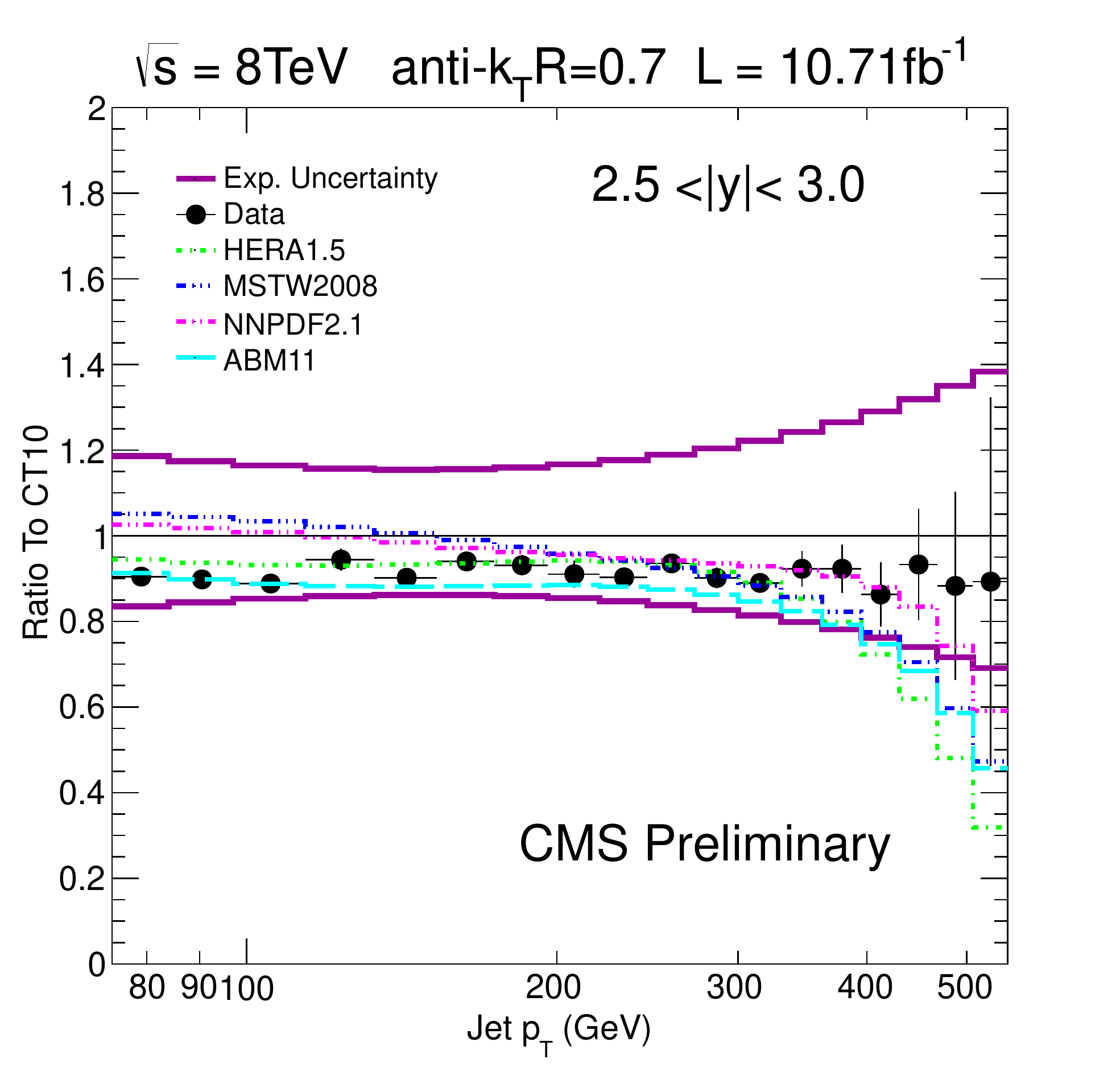}
\caption{The ratio of measured inclusive jet cross-section at COM energy $\rm \sqrt{s}~ =~ 8~ TeV$ to the CT10 theory prediction is shown along with the ratio to other PDF sets and relative
 total experimental uncertainties for the two extreme $\rm |y|$ bin. For all $\rm |y|$ bins a good agreement between data and CT10 theory is observed.}
\label{fig:ratio8}
\end{figure}

An analogous inclusive jet cross-section measurement is carried out with $\rm \sqrt{s}~ =~ 8~ TeV$ ($\rm \mathcal{L}_{int}~=~10.71~fb^{-1}$) \cite{JET8} CMS data. In this analysis the measurement
 extends in rapidity upto $\rm |y|~=~3.0$ and the jet $\rm p_{T}~=~2.5~TeV$. For the theory comparison ABM11 PDF set is used instead of ABKM09. The six single jet HLT triggers
used to reconstruct the spectrum have the threshold values
$\rm p_{T}>$ 40 GeV, 80 GeV, 140 GeV, 200 GeV, 240 GeV and 320 GeV respectively. As seen in Fig.\ref{fig:ratio8}, for the central region, data is within 5$\%$ agreement to CT10 theory prediction whereas
it fluctuates up to 10$\%$ for the outer most bin. The total experimental uncertainty in central region varies within 5-20$\%$ but reaches to 40$\%$ for the outermost rapidity bin.  In the outer rapidity
bin, for the high $\rm p_{T}$ region, the theory predictions mismatch among themselves significantly.

\section{Cross-section ratio measurements}
The collinear branching of partons in a QCD process carry away the momenta carried by the original mother parton. In order to reconstruct the four momenta of the original hard scattered partons,
it is essential to cluster all these radiations in jet clustering and hence a wider jet radius is preferable. At the same time jets with larger radii gets more contribution from pileup, distorting
their four momenta distribution. So an optimal intermediate value of the jet radius is preferable in order to balance between these two effects.

\begin{figure}
\hspace{0.5cm}
\includegraphics[scale=0.27]{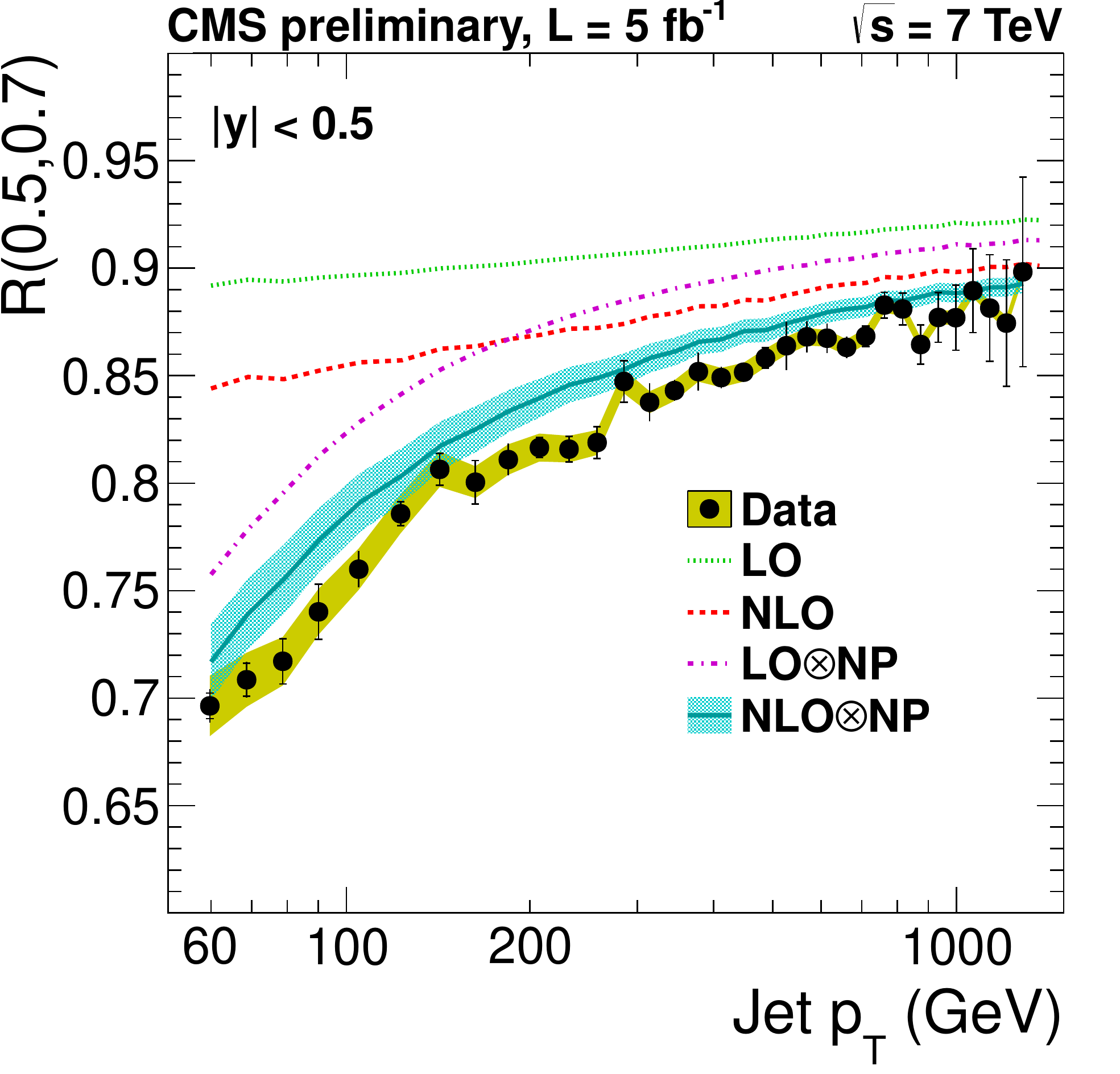} \hspace{1.5cm}
\includegraphics[scale=0.27]{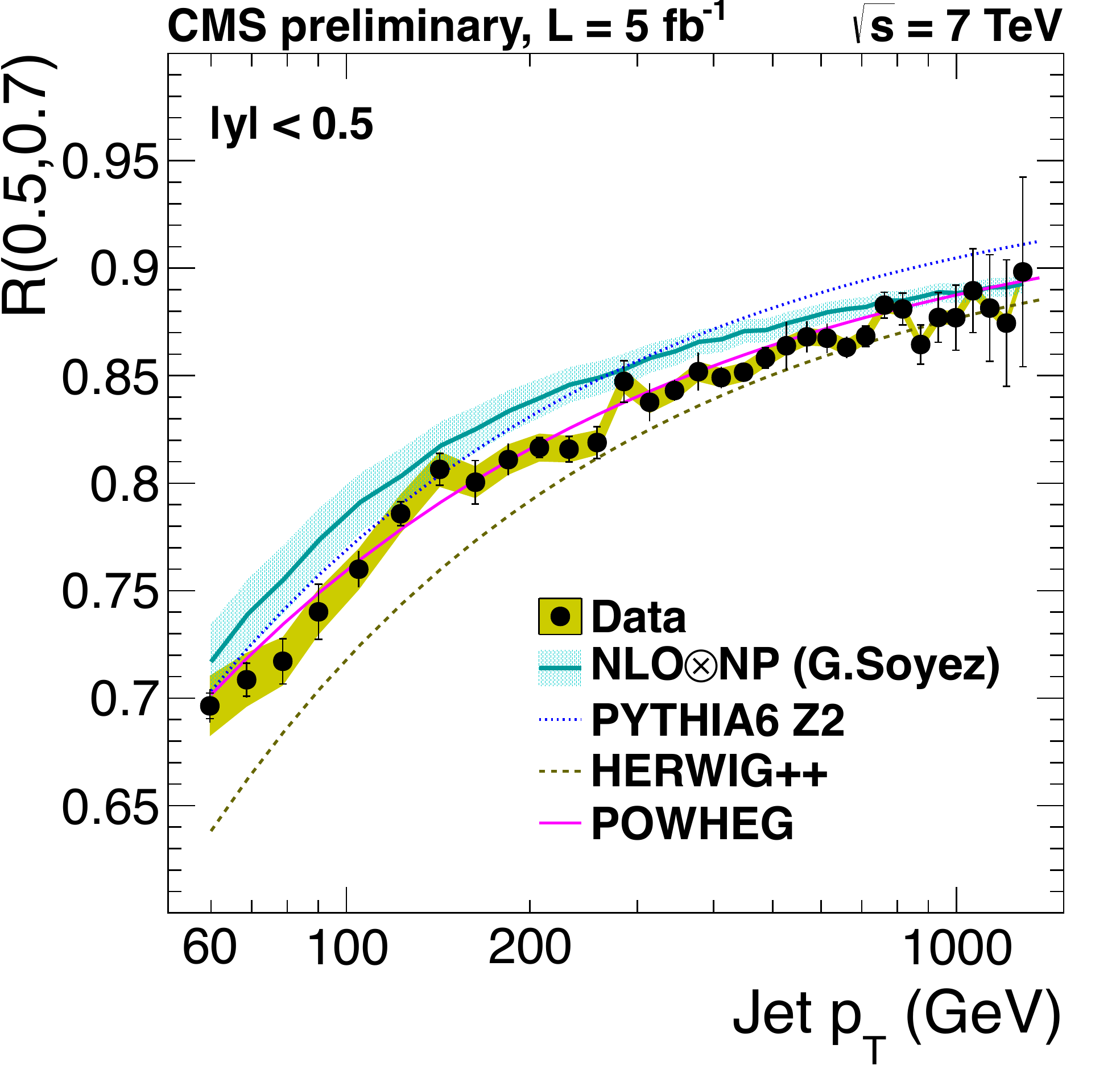}
\caption{The left plot shows the $\rm R(0.5,0.7)$ for data with total experimental uncertainty as a function of jet $\rm p_{T}$
 compared with LO, NLO, $\rm LO\otimes NP$ and $\rm NLO\otimes NP$ theory prediction. The right plot shows a comparison between data and different generator predictions. 
}
\label{fig:ak5ak7}
\end{figure}

In order to optimize jet radius, a measurement of inclusive jet cross section ratio (R(0.5,0.7)) is carried out in CMS where the jets are reconstructed with
 AK5 and AK7 algorithm \cite{AK5AK7}. The dataset used is 
$\rm \sqrt{s}~ =~ 7~ TeV$ ($\rm \mathcal{L}_{int}~=~5~fb^{-1}$) CMS data.
 In this case the jet $\rm p_{T}$ measurement goes down to 60 GeV. In Fig.\ref{fig:ak5ak7} the comparison between measured
 ratio to different order theory calculations of R(0.5,0.7) are shown. The $\rm NLO\otimes NP$ theory prediction matches with data well compared
 to $\rm LO\otimes NP$ predictions. The pQCD predictions without NP corrections
are in clear disagreement with the data. A feature of the ratio R(0.5,0.7) is that the ratio doesn't change significantly for a given $\rm p_{T}$ bin across the rapidity range $\rm |y|<2.5$.
 The comparison of data with theories
 from different generator prediction shows that POWHEG matches with data best in most of the $\rm p_{T}$ region. The fixed order pQCD calculation matched with parton shower has in general a better
agreement with data compared to pQCD calculations corrected for the NP effects. In this measurement pileup is the main source of experimental uncertainty as two jets with different radii have different
amount of pileup clustering.

This measurement shows that collinear radiation is one of the main source of discrepancy between measurement and pQCD theory predictions. Hence larger radius jets are more reliable for absolute 
cross-section measurements and PDF fits. The optimal radius is a trade-off between collinear radiation($\rm \sim |logR|$), NP effects($\rm \sim  1/R$) and pileup contribution($\rm \sim R^{2}$).

\section{Color coherence measurement}
In a parton scattering process, the outgoing color connected final state partons keep on interfering among themselves while fragmentation and this phenomena is known as color coherence. 
The angulur correlation between 2nd and 3rd jet in an event (ordered by $\rm p_{T}$) is measured with CMS $\rm \sqrt{s}~ =~ 7~ TeV$ ($\rm \mathcal{L}_{int}~=~36~pb^{-1}$) data \cite{COLCOH}.
 The variable designed
to study the color coherence effect is $\rm \beta=tan^{-1}[(\phi_{3}-\phi_{2}),sgn(\eta_{2}).(\eta_{3}-\eta_{2})]$ . In presence of color coherence, an enhancement of 3rd jet population is expected
in the region $\rm \beta~=~0~ or ~ \pi$ and a suppression is expected in the region $\rm \beta~=~\pi/2$. The events are selected with at-least 3 jet whose $\rm p_{T}>30~GeV$ and $\eta_1,\eta_2<2.5$.

\begin{figure}
\includegraphics[scale=0.38]{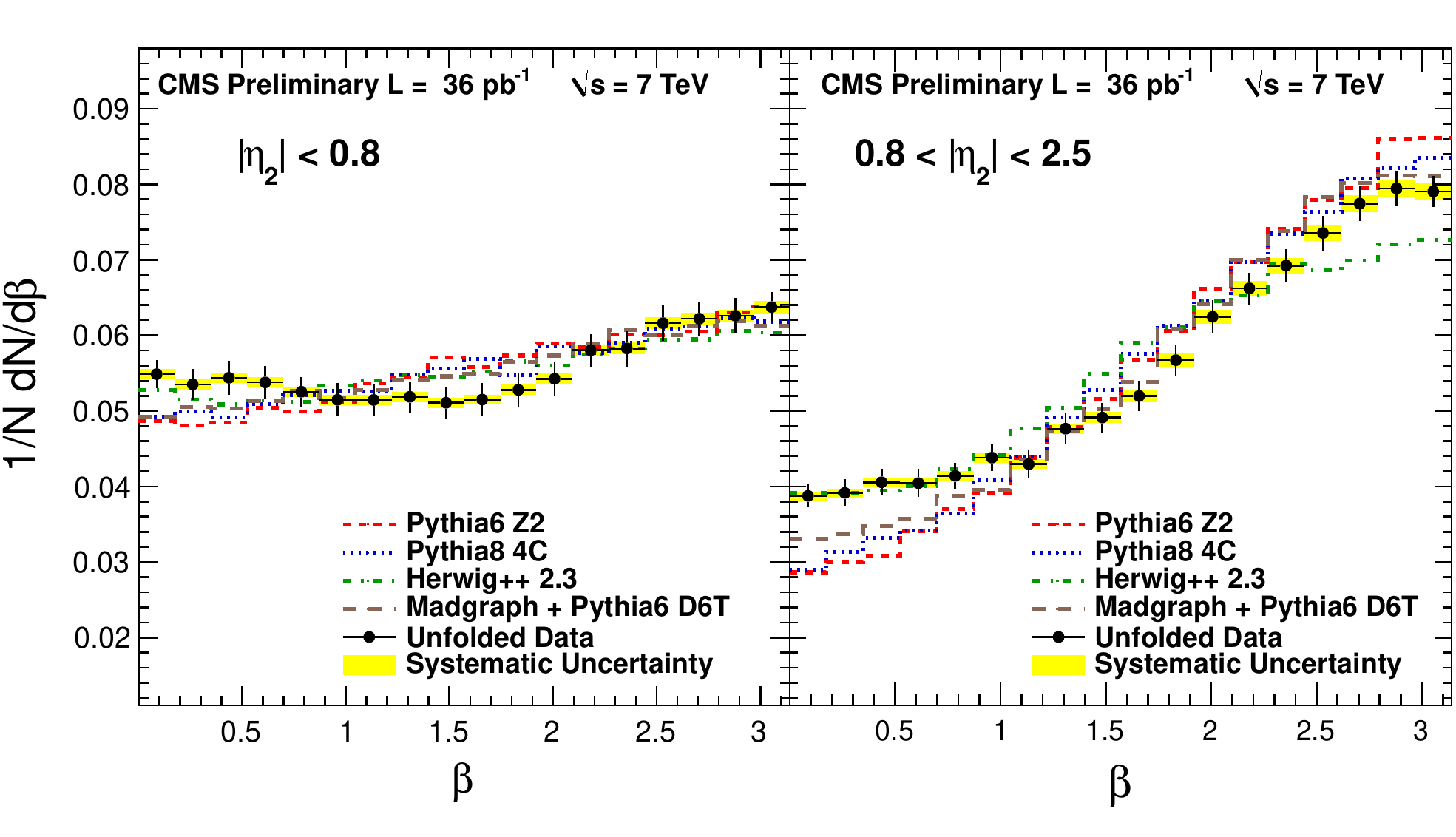}
\includegraphics[scale=0.38]{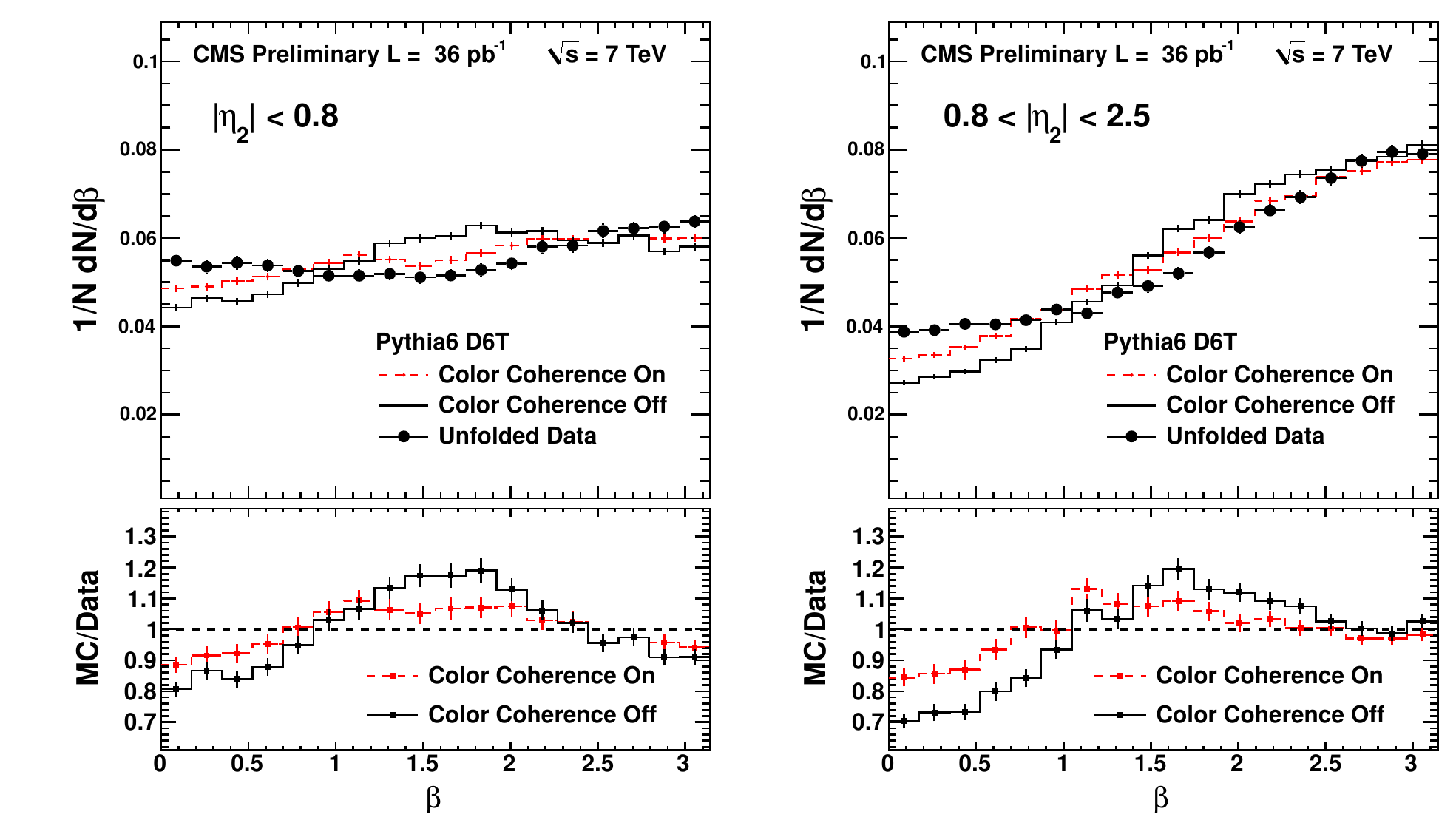}
\caption{The left plot shows the distribution of $\beta$ in data and four different MC for two different $\eta_2$ regions. The last two plots show that effect of switching on the color coherence in MC 
gives a better agreement with data.}
\label{fig:colcoh}
\end{figure}

In Fig.\ref{fig:colcoh} we see that the distribution of $\beta$, compared between data and different MC predictions. In general HERWIG++ shows a better agreement with data except in the forward region
$\beta \sim \pi$. The middle and the right plot in the same figure shows that MC predictions with color coherence effect switched on gives better agreement with data compared to the MC distributions with 
color coherence turned off. This study verifies the color correlation among outgoing partons in a hard scattering process.

\section{Summary}
In this proceedings, first we discussed about PF algorithm and jet reconstruction in CMS. The multiplicative jet energy correction procedure is described followed by a discussion on evaluation of JES
uncertainties from different sources as evaluated from the full $\rm \sqrt{s}~ =~8~TeV$ CMS data. 
Next we discuss about the absolute jet production cross-section measurements at CMS for both $\rm \sqrt{s}~ =~ 7~ TeV,~8~TeV$ datasets.
 Double differential inclusive and dijet production cross-section measurements are reported along with the comparison
to $\rm NLO \otimes NP$ theory predictions from five different PDF sets.   

Cross-section ratios of two different jet radii were reported following the discussion on absolute cross-section measurements.
 A comparison between data and different MC generators
 show that fixed order pQCD prediction matched with parton shower in general agrees with data well.
This results also optimize the jet radius to be fixed in order to incorporate the collinear branching of partons, NP effects and pileup effect. 

Finally we discussed the study of color-coherence effects in the fragmentation of final state partons for three-jet final states.
 Color-coherence effects were demonstrated by switching it on and off in PYTHIA6 spectrum and comparing it with the measured observable
which clearly showed a favour towards data with color-coherence effects turned on.

In general jet physics study is a very vibrant activity within CMS. An excellent jet performance is obtained for the two LHC runs.
 Starting from jet object reconstruction to several different measurements are carried out within CMS. Some more results of extracting
$\alpha_S$ running and PDF fitting with the measured jet cross-sections are coming up. Overall the jet studies at CMS are leading towards better understanding of QCD in new energy regime.

\end{document}